# Observation of thermal acoustic modes of a droplet coupled to an optomechanical sensor

G. J. Hornig, K. G. Scheuer, and R. G. DeCorby[a]

[1] *ECE Department, University of Alberta, 9211-116 St. NW, Edmonton, AB, Canada, T6G 1H9*

The bulk acoustic modes of liquid droplets, well understood from a theoretical perspective, have rarely been observed experimentally. Here, we report the direct observation of acoustic vibrational modes in a picoliter-scale droplet, extending up to ~ 40 MHz. This was achieved by coupling the droplet to an ultra-sensitive optomechanical sensor, which operates in a thermal-noise limited regime and with a substantial contribution from acoustic noise in the ambient medium. The droplet vibrational modes manifest as Fano resonances in the thermal noise spectrum of the sensor. This is amongst the few reported observations of droplet acoustic modes, and of Fano interactions in a coupled mechanical oscillator system driven only by thermal Brownian motion.

___________________________

[a] Author to whom correspondence should be addressed. Electronic mail: rdecorby@ualberta.ca

The vibrational modes of liquid droplets have been studied for well over 100 years.[1] These modes essentially fall into two categories[2]: i. shape oscillations for which surface tension acts as the restoring force (i.e. capillary waves), and ii. 'bulk' acoustic resonances inside the droplet mediated by compression and rarefication of the liquid medium. A vast body of literature exists on the study of droplet capillary oscillations,[3-5] which typically have fundamental resonance frequencies in the kHz range. Conversely, the fundamental acoustic resonances of small droplets lie in the MHz range and have rarely been accessible from an experimental perspective. Dahan *et al.*[2] observed the fundamental acoustic mode (and associated harmonics) for liquid droplets driven by resonantly circulating laser light (i.e. acting simultaneously as optical and mechanical resonators). A similar setup was used to excite surface acoustic waves in the MHz range through stimulated Brillouin scattering.[7] To our knowledge, passive observation of the acoustic modes of a liquid droplet driven only by thermal Brownian motion has not been previously reported.

To appreciate the challenge with 'passively' observing droplet acoustic modes, note that the mean-square pressure generated by Brownian motion of the molecules in a liquid medium can be estimated (from equipartition theory [8]) as $<p^2> = 4\pi kT\rho f^2 \Delta f/v$, where $\rho$ is the medium density, $v$ is the sound velocity, and $\Delta f$ is a frequency interval. Using $p = 2\pi f \rho v A$, and considering a water medium at 300 K as a typical example, this corresponds to RMS displacement spectral density $A_{RMS} \sim 2\times10^{-17}$ m/Hz$^{1/2}$. Remarkably, optomechanical sensors[9-11] have achieved displacement sensitivities at least 1-2 orders of magnitude lower than this. Moreover, the on-resonance displacement for acoustic modes in a droplet (Q ~ 100,[2]) is expected to be ~ 1 order of magnitude higher than the mean displacement associated with the thermal noise of the fluid medium.

Viewed from another perspective, the pressure equivalent of the thermal noise from the fluid medium for an acoustic (i.e. ultrasound) sensor of area $S$ can be expressed[12] $NEP_M = (kT\rho v/S)^{1/2}$, which sets a lower limit on the sensitivity of any ultrasound sensor operating in a given medium.[13] For example, we recently[14] described optomechanical ultrasound sensors with $S \sim 10^{-8}$ m$^2$ corresponding to $NEP_M \sim 0.8$ mPa/Hz$^{1/2}$ in room-temperature water. Notably, our experimentally estimated $NEP$ (in the $\sim$ 0-15 MHz range) lies very close to this value, indicating that the thermal noise from the surrounding fluid medium is the dominant contributor to the noise spectrum.

Of related interest, it has been known for some time that the noise signal of an acoustic/ultrasound sensor contains information about its environment.[15] At ultrasound frequencies, the initial experiments employed piezoelectric sensors,[16] for which the noise signal is substantially attributable to electrical noise and intrinsic thermal noise. This necessitates extensive averaging and signal processing to extract information from the random thermal noise. Moreover, piezo-based sensors must have a large area in order to approach the required pressure sensitivities,[12] which limits the frequency range and resolution for ultrasound imaging. Nevertheless, environmental information was extracted from thermal acoustic noise (with displacement spectral density $\sim 3 \times 10^{-18}$ m/Hz$^{1/2}$) inside an aluminum block.[16] More recently, capacitive membrane (CMUT) sensors were used to perform similar experiments in water.[17]

Since optomechanical sensors can readily operate in a thermal medium-noise dominated regime, and can achieve these sensitivities with micron-scale dimensions,[13,14,18] they might enable powerful options for passive imaging of small structures. Here, we demonstrate the observation of acoustic vibrational signatures for a droplet coupled to an optomechanical sensor. This can be viewed as passive sensing of the acoustic environment by the optomechanical device, or alternatively as coupling[19] between the vibrational modes of the droplet and the optomechanical device.

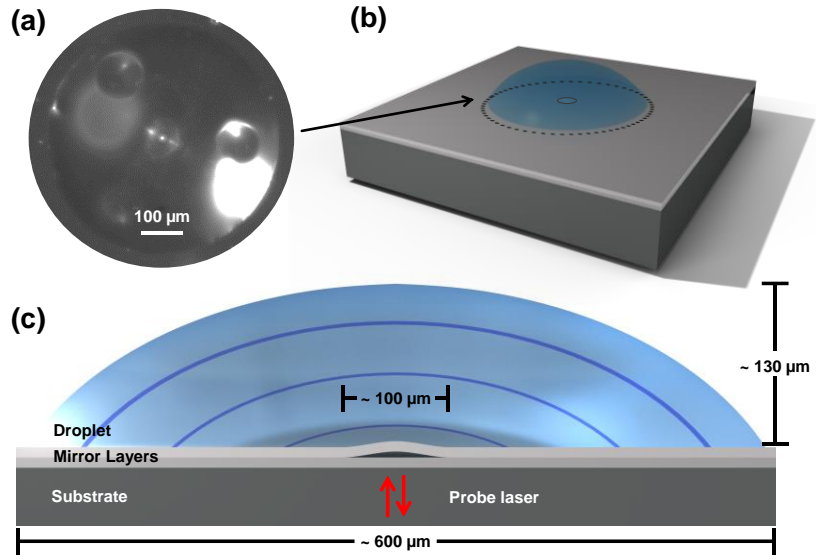

FIG. 1. (a) Photograph of droplet aligned overtop an array of four ultrasound sensors, and concentric with the main device of interest here. (b) Artists depiction of the droplet placed on the detector and surrounding substrate, and (c) cross-sectional depiction of the droplet aligned overtop a microcavity ultrasound sensor. Thermal Brownian motion from the fluid is resonant within the droplet, resulting in the emergence of high-amplitude acoustic features (depicted by blue concentric rings.)

For the present study, we utilized optomechanical ultrasound sensors[14] fabricated in a customized thin-film buckling process. These devices are plano-concave optical microcavities with a sealed and partially evacuated cavity enclosed by thin film dielectric mirrors. The curved (buckled) upper mirror also functions as a mechanical (membrane-like) resonator, with its low-order vibrational resonance frequencies in the MHz range.[20] Time-varying pressure waves deflect this membrane mirror, and the change in cavity height is interrogated by a low-power (<< 1 mW) infrared laser incident from the opposite side. Using a tuned-to-slope technique, acoustic (ultrasound) signals are transduced to variations in the power of the reflected laser signal. It is important to note that the laser probe does not significantly drive any of the processes described below, for example by heating of the membrane or droplet, which we confirmed by monitoring the position of the optical resonance.[20] Moreover, we confirmed that the spectral features described below were insensitive to coupled laser power (typically varied in the 10s to 100s of µW range) or to operating on the red- or blue-detuned side of the optical resonance. This implies that the measurements are not significantly impacted by 'optomechanical' (i.e. dynamical back-action[9]) effects driven either by thermal or radiation pressure effects. Thus, the laser simply facilitates the passive interrogation of mechanical motion[11] in our experiment.

Our sensors combine state-of-the art pressure sensitivity ($NEP < 1$ mPa/Hz$^{1/2}$ in water) with small footprint (which also makes their response nearly omnidirectional), and their sensitivity is limited by thermo-mechanical noise (mainly thermal medium noise, as mentioned above) over a broad range of frequencies. The devices here are similar to those described in our previous work,[14] the primary difference being fabrication on a fused silica substrate instead of silicon. To facilitate the placement of droplets, patterned regions of lower and higher hydrophobicity were added onto the chips used here, by lifting off a ~ 200 nm thick fluorocarbon layer deposited via CVD on their top silicon surface. Specifically, most of the surface was made hydrophobic except for circular patterns (which we refer to as hydrophilic regions) centered on groups of the devices. This allowed for sessile droplets with a circular base to be controllably centered on an ultrasonic sensor. Here, we focus our discussion on the particular geometry shown in Fig. 1, where a hydrophilic region 600 μm in diameter is nominally centered on an array of four sensors – one in the center, and three placed symmetrically 120 degrees apart in a concentric circular pattern 300 μm in diameter.

Droplets were formed by placing a small amount of glycerol on a chip, and then spreading it laterally into a thin layer using a soft stick (the 'squeegee' method.) The thin layer quickly evacuated regions coated with the hydrophobic fluorocarbon, leaving behind small circular drops that conformed closely to the patterned 'holes' in the fluorocarbon layer. From side-view camera images (not shown), the droplets studied here were estimated to have a peak height ~ 130 μm and a near-ideal hemispherical-cap geometry, thus corresponding to a volume of approximately 125 pL.

A typical spectrum (in the 0-40 MHz range) recorded by an optomechanical sensor centered on the droplet is shown in Fig. 2. The blue trace was captured with the interrogation laser tuned near an optical cavity resonance, and represents the thermomechanical noise of the optomechanical sensor. The dominant features (~ 6-7 broad peaks) in the spectrum correspond to damped vibrational modes of the buckled mirror in the glycerol medium,[14] and we have confirmed that these features do not change significantly as more liquid is added onto the sensor. The black trace was captured with the laser detuned from the cavity resonance and with the same average receive power at the photodetector, and thus represents the shot noise floor of the system. The optomechanical sensor has a resonant characteristic, with a response that varies by 1-2 orders of magnitude. However, because the noise floor is dominated by thermal medium noise (see above), the device is highly sensitive to external signals over the entire frequency range, and in fact the NEP of such a sensor is essentially frequency independent (i.e. both the response and the noise are enhanced by the same amount near a resonance).[14,18]

With the adjacent glycerol medium in the form of a small droplet having the dimensions described above, we observed the emergence of new 'Fano-like' features within the thermomechanical noise spectrum. It is worth emphasizing that these features were observed only when the adjacent glycerol medium was formed into a small droplet. Moreover, amongst many measurements (see supplementary material), the lowest-frequency 'Fano-like' feature was consistently located at ~ 5.4 MHz, as indicated for example in Fig. 2. Other similar features, of varying strength, were observed at higher frequencies. For the case shown in Fig. 2, strong features can also be seen at ~ 10.3 MHz and ~ 28 MHz, while weaker features can be discerned throughout the measurement range (for example, at ~11, 11.5, 12, and 34 MHz). Data for other sensor/droplet combinationsis provided in the supplementary material.

The asymmetric line-shape of these features is distinctly characteristic of a so-called Fano resonance [19,21-23], which arises from the interaction between a localized (i.e. resonant) mode and a continuum of background modes. A system of two coupled harmonic oscillators is widely used[22] to explain the underlying physics, but only recently[23] has Fano interference been experimentally verified for a purely mechanical system. Here, as shown below, the relatively high-Q acoustic modes of the droplet act as the discrete states, while the highly damped thermo-mechanical vibration of the optomechanical resonator acts as the background continuum. Unlike the actively driven mechanical oscillators described by Stassi et al.,[23] here both sets of modes are driven only by thermal Brownian motion.

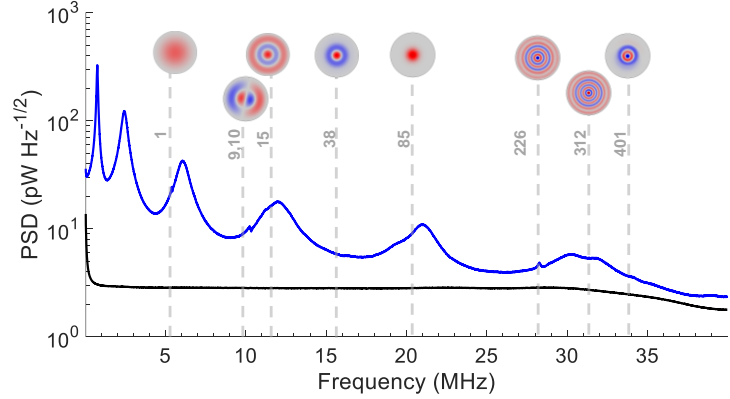

FIG. 2. Power spectral density of the noise signal for the central detector shown in Fig. 1a (blue,) alongside the shot noise floor (black.). The numerically predicted eigenfrequencies for several acoustic modes of the droplet are indicated by the vertical dashed lines. The corresponding pressure distributions (at the droplet-substrate interface) are also shown, with red/blue indicating pressure standing-wave antinodes of opposite phase. See supplementary material for further information.

We modeled the observed coupling using a thermal Langevin force driven linear coupled oscillator model adapted from Lin et al.[24], which describes interactions between a 'bright' and a 'dark' mechanical mode (i.e. strongly and weakly coupled to the interrogation laser light field, respectively). In our system, the freely moving mirror of the optomechanical sensor represents the 'bright' mode, while the acoustic vibration of the adjacent droplet (which is coupled to the light field only indirectly through mechanical interaction with the membrane mirror) represents the 'dark' mode. In this model, we used the observed center frequency and quality factor of the optomechanical membrane mode nearest to the Fano feature of interest, and fit the observed noise spectrum to extract the same parameters for a second damped oscillator, in addition to an elastic coupling parameter linking the two oscillators, and an effective mechanical mass (see the supplementary material for the equations used, the fitting parameters extracted, and a more detailed explanation).

Sample fits for the two lowest-frequency Fano-like features are shown in Fig. 3 (dashed red and green respectively), revealing good agreement between experiment and theory. The extracted spectral profile for the second oscillator (i.e. the 'dark' mode attributed to the droplet,) is overlaid with an arbitrary amplitude as a reference (dotted, colored respectively). For these two modes, we estimated a resonant frequency of ~ 5.41 MHz and 10.25 MHz, and a Q-factor of 164 and 73, respectively. Notably, these Q-factors are in-line with theoretical predictions based on viscosity-damping for a droplet of this size.[2] Moreover, the extracted coupling coefficients (~ 1 MHz) are very close to the overall damping rate extracted for the associated membrane mode in each case, as would be expected if this damping is indeed dominated by coupling to the glycerol droplet. Finally, very reasonable values for the effective masses of the mirror vibrational modes (see the supplementary material file) were also extracted. These lend strong support to our assertion that the observed Fano-like features are in fact signatures of the droplet acoustic modes.

To further corroborate our theory regarding the origin of these Fano features, we modeled the droplets as acoustic cavities, which possess eigenfrequencies dependent on the geometry and the speed of sound in the fluid. We numerically simulated these eigenfrequencies using a linear elastic fluid model (see supplementary material for details,) in which a set of eigenmodes in close analogy to the acoustic modes of a perfect hemisphere are predicted.[25] Selected eigenfrequencies predicted for a droplet 130 μm tall and 600 μm in diameter are shown overlaid in Fig. 2 and Fig. 3 as vertical dashed grey lines, with corresponding images of the pressure distributions at the substrate surface as insets in Fig. 2. The agreement between the simulated lowest-frequency eigenmode (at ~ 5.36 MHz in Fig2(a)) and the fitted oscillator extracted from the strongest Fano feature (at ~ 5.41 MHz) is excellent, and this feature was consistently observed for multiple droplet/sensor combinations (see supplementary material). More speculatively, we attribute the second-strongest Fano feature in Fig. 3 (at ~ 10.25 MHz) to the ninth (degenerate) simulated acoustic mode at 9.9 MHz (Fig. 2(a)).

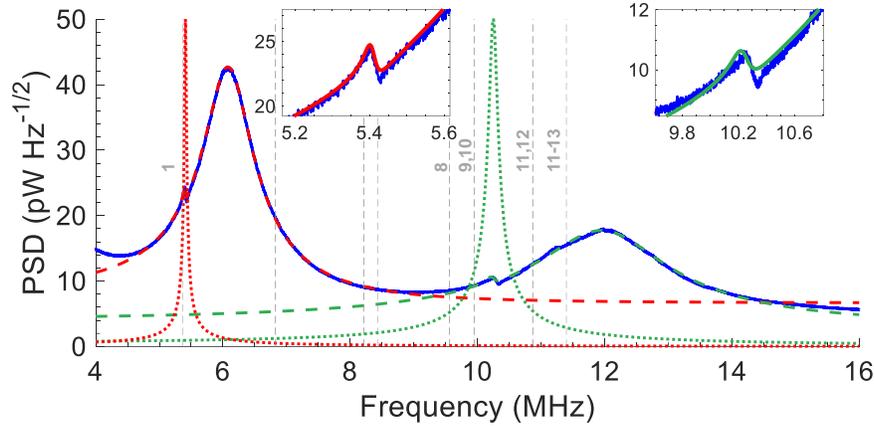

FIG. 3. Fits of analytical coupled-resonator theory for the two most prominent Fano features visible (at ~ 5.4 MHz and ~10.3 MHz in dashed green and red respectively,) using parameters detailed in the Table S1 of the supplementary material. Analytically predicted droplet line-shapes for each mode are also overlaid (dotted, colored respectively,) plotted with an arbitrary amplitude. Numerically predicted modes are shown as vertical dashed lines (see Fig. 1.)

For a circular droplet and a central detector, one might expect that modes with central pressure anti-nodes would provide the strongest signal. Indeed, this is particularly the case for the lowest frequency mode at ~ 5.4 MHz, which as mentioned was consistently observed across multiple droplet/device measurements. However, the strength of the interaction between a particular pair of modes is a complex problem for the higher-order modes. The higher-order mechanical modes of the optomechanical membrane themselves have complicated spatial patterns (similar to those of the droplet) and it is the spatial overlap between the nodes of these membrane-mechanical and droplet-acoustic modes that impacts the likelihood of detecting their interaction. For example, we measured evidence of coupling to different droplet modes for non-central sensors within the same droplet (such as the one shown on the right in Fig. 1(a), see supplementary material.) Furthermore, even droplet modes exhibiting a central pressure node might still impart significant pressure oscillations across the footprint of the optomechanical detector, whose base diameter is ~ 20% of the droplet diameter (i.e. it cannot be viewed as a 'point' detector). Moreover, the nature and strength of a Fano interaction is dependent on other details besides the coupling strength, including the frequency spacing and the relative damping of the two modes.[21-23] A more complete exploration of these details is ongoing, but it is worth noting that in experiments to date we have found that mutual centering of the droplet and the optomechanical sensor is favorable for observing strong coupling features.

In summary, we have reported passive observations of the bulk acoustic modes of a droplet via their coupling to an optomechanical sensor. Previous schemes used to measure droplet bulk acoustics have only recently been reported, and used some form of drive or pump to excite these modes.[2] Similarly, the recently reported[23] observation of Fano resonance in a system of coupled mechanical oscillators required active driving of the resonators by a piezo actuator. The ability to observe these features in such a simple system as described here, and moreover under the action of only intrinsic Brownian motion, is remarkable evidence of the extreme sensitivity made possible by optomechanical sensors.[11]

Aside from their theoretical interest, these results have potential applications. For example, there is a long-standing interest in ultrasonic probing of small-volume liquid samples for molecular physics and biology.[26] The information contained in the passive acoustic resonances of a droplet might be used to directly sense the geometrical (shape, surface tension, etc.) or chemical (density, speed of sound, viscosity etc.) properties of the fluidic droplet itself. Sessile-droplet-based 'open' microfluidics platforms are in fact a topic of active development,[27] but so far rely mainly on electrical or optical sensing modalities. Our devices could enable new options for 'acoustic spectroscopy' at the microscopic scale. We hope to explore these possibilities in future work.

**SUPPLEMENTARY MATERIAL**

See supplementary material for additional information on the coupled-oscillator theory, fitting parameters for the figure shown here, and additional corroborating measurements.

[1] Lord Rayleigh, Proc. R. Soc. Lond. A 29 71-97 (1879.)
[2] R. Dahan, L. L. Martin, and T. Carmon, Optica 3, 2, 175 (2016).
[3] J. Blamey, L. Y. Yeo, and J. R. Friend, Langmuir 29, 3835 (2013).
[4] S. Maayani, L. L. Martin, S. Kaminski, and T. Carmon, Optica 3, 5, 552 (2016).
[5] X. Luo, Z. Zhou, W. Liu, D. Shen, H. Yan, Y. Lin, and W. Wan, Opt. Lett. 46, 18, 4602 (2021).
[6] S. Sharma and D. I. Wilson, J. Fluid Mech. 919, A39 (2021).
[7] A. Giorgini, S. Avino, P. Malara, P. De Natale, and G. Gagliardi, Opt. Lett. 43, 15, 3473 (2018).
[8] R. H. Mellen, J. Acoust. Soc. Am. 24, 5, 478 (1952).
[9] A. Schliesser, G. Anetsberger, R. Riviere, O. Arzicet, and T. J. Kippenberg, New J. Phys. 10, 095015 (2008).
[10] R. Singh and T. P. Purdy, Phys. Rev. Lett. 125, 120603 (2020).
[11] B.-B. Li, L. Ou, Y. Lei, and Y.-C. Liu, Nanophotonics 10, 11, 2799 (2021).
[12] A. M. Winkler, K. Maslov, and L. V. Wang, J. Biomed. Opt. 18, 9, 097003 (2013).
[13] S. Basiri-Esfahani, A. Armin, S. Forstner, and W. P. Bowen, Nat. Commun. 10, 1, 132 (2019).
[14] G. J. Hornig, K. G. Scheuer, E. B. Dew, R. Zemp, and R. G. DeCorby, Opt. Express 30, 18, 33083 (2022).
[15] R. L. Weaver, Science 307, 1568 (2005).
[16] R. L. Weaver and O. I. Lobkis, Phys. Rev. Lett. 87, 13, 134301 (2001).
[17] S. Lani, S. Satir, G. Gurun, K. G. Sabra, and F. L. Degertekin, Appl. Phys. Lett. 99, 224103 (2011).
[18] W. J. Westerveld, Md. Mahud-Ul-Hasan, R. Shnaiderman, V. Ntziachristos, X. Rottenberg, S. Severi, and V. Rochus, Nature Photon. 15, 341 (2021).
[19] M. H. J. de Jong, J. Li, C. Gartner, R. A. Norte, and S. Groblacher, Optica 9, 2, 170 (2022).
[20] M. H. Bitarafan, H. Ramp, T. W. Allen, C. Potts, X. Rojas, A. J. R. MacDonald, J. P. Davis, and R. G. DeCorby, J. Opt. Soc. Am. B 32, 1214 (2015)
[21] U. Fano. Phys. Rev. 124, 1866 (1961).
[22] A. E. Miroschnichenko, S. Flach, and Y. S. Kivshar, Rev. Mod. Phys. 82, 3, 2257 (2010).
[23] S. Stassi, A. Chiadò, G. Calafiore, G. Palmara, S. Cabrini, and C. Ricciardi, Sci. Rep. 7, 1065 (2017)
[24] Q. Lin, J. Rosenberg, D. Chang, R. Camacho, M. Eichenfield, K. J. Vahala, and O. Painter, Nature Photon 4, 236 (2010).
[25] J. L. Flanagan, J. Acoust. Soc. Am. 37, 4, 616 (1965)
[26] A. P. Sarvazyan, Ultrasonics 20, 4, 151 (1982).
[27] J. L. Garcia-Cordero and Z. H. Fan, Lab Chip 17, 2150 (2017).